\documentclass[aps, prd, onecolumn, 
showpacs, showkeys, 
print,
groupedaddress,
nofootinbib
]{revtex4-2}

\usepackage[utf8]{inputenc}
\usepackage[T1]{fontenc}
\usepackage{lmodern}
\usepackage{microtype}
\usepackage[english]{babel}

\usepackage{amssymb,amsbsy,amsmath,amsfonts,amsthm}
\usepackage{yhmath} 

\usepackage{slashed}
\usepackage{bm}
\usepackage{mathptmx}  % replace package times

\usepackage[dvipsnames,table,xcdraw]{xcolor}
\definecolor{lightyellow}{RGB}{255,250,205}

\usepackage{graphicx}
\usepackage{epstopdf}

\usepackage{array}        % array first 
\usepackage{tabularx}
\usepackage{multirow}
\usepackage{longtable}

\usepackage{url}
\usepackage{orcidlink}

\usepackage{hyperref} % hyperref puts last!
\hypersetup{ %fill inside borders: %false: boxed links; true: colored links
  colorlinks=true,
  linkcolor=blue,  
  citecolor=cyan,
  urlcolor=red,           
}

\newcommand{\tch}{t_{0,\mathrm{ch}}}
\newcommand{\tstar}{t_0^{*}}
\newcommand{\ktt}{\tilde k_1}

\begin{document}

\title{Gradient-flow scale setting from octet baryon masses in N$^3$LO covariant chiral perturbation theory}

\author{
Xiu-Lei Ren\orcidlink{0000-0002-5138-7415}}
\email[]{xiulei.ren@sdu.edu.cn}
\affiliation{Shandong Provincial Key Laboratory of Nuclear Science, Nuclear Energy Technology and Comprehensive Utilization, 
\& School of Nuclear Science, Energy and Power Engineering, Shandong University, 
250061 Jinan, China}

\begin{abstract}
We determine the gradient-flow scale $t_0$ by analyzing the lowest-lying octet-baryon masses on the Coordinated Lattice Simulations (CLS) $N_f=2+1$ ensembles in covariant baryon chiral
perturbation theory up to next-to-next-to-next-to leading order (N$^3$LO). By including the finite-volume corrections self-consistently 
and the discretization effects from Symanzik effective field theory, we obtain a good description of the RQCD data and determine the gradient-flow scale to be $\sqrt{8t_{0,\mathrm{ph}}}=0.4120(14)_\mathrm{stat.}(16)_\mathrm{sys.}$~fm with sub-percent precision. The predicted octet baryon masses agree well with the experimental values in the isospin limit. Using the Feynman-Hellmann theorem, we also estimate the nucleon sigma terms, which are consistent with direct determinations from three-point correlation functions on the same CLS ensembles.
\end{abstract}

%\pacs{14.40.Nd, 14.40Rt, 11.80.Jy}
%\keywords{}
%\preprint{arXiv: xxxxxx }

\maketitle

\date{\today}

\section{Introduction}

Lattice QCD (LQCD) has become an indispensable tool for investigating the strong interaction phenomena from first principles \cite{Rothe:1992nt,Gattringer:2010zz}. 
Since lattice simulations can only give the dimensionless physical quantities, such as hadron masses in units of the lattice spacing $a$, expressing any dimensionful observable in physical units requires an external physical input to fix $a$. This procedure is called scale setting. There are two classes of quantities commonly used for scale setting in LQCD~\cite{FlavourLatticeAveragingGroupFLAG:2024oxs}: the physical scales, e.g., the masses of stable baryons such as $\Omega$, the leptonic decay constants of the pion and kaon mesons; and the theoretical (intermediate) scales, e.g., the quark-antiquark potential scales $r_0$~\cite{Sommer:1993ce} and $r_1$~\cite{Bernard:2000gd} and the gradient-flow scales $t_0$~\cite{Luscher:2010iy} and $w_0$~\cite{BMW:2012hcm}. 
For the intermediate scales, a physical observable is still required to determine its value in physical units. Note that, in both cases, the scale is ultimately obtained from lattice data extrapolated to the physical quark masses, the continuum and infinite-volume limits. 

Among the theoretical scales, the gradient-flow scales have become particularly attractive because of their favourable statistical properties~~\cite{Luscher:2010iy,BMW:2012hcm}. The scales $t_0$ and $w_0$  are defined  in terms of the gluonic action density at positive flow time and its flow-time derivative, respectively. Thus their determination does not need to fit the long-distance correlation functions and can achieve high statistical precision. 
Furthermore, the quark-mass dependence of $t_0$ has been derived in chiral perturbation theory (ChPT) with nonanalytic contributions first appearing at next-to-next-to-leading order (NNLO)~\cite{Bar:2013ora}. 
The current status of $t_0$ is summarized in Sec.~11 of the FLAG24 review \cite{FlavourLatticeAveragingGroupFLAG:2024oxs}, and the averaged value $\sqrt{t_0}=0.14474(57)$ fm for $N_f=2+1$ is obtained based on four  lattice simulations by RQCD~\cite{RQCD:2022xux}, CLS~\cite{Bruno:2016plf}, RBC/UKQCD~\cite{RBC:2014ntl} and BMW~\cite{BMW:2012hcm} collabrations. Compared with the previous FLAG21 average value $\sqrt{t_0}=0.14464(87)$ fm~\cite{FlavourLatticeAveragingGroupFLAG:2021npn}, the inclusion of the new RQCD result~\cite{RQCD:2022xux} has reduced the uncertainty from about $0.6\%$ to $0.4\%$.

The RQCD determination is based on an extensive set of octet-baryon mass data from the CLS ensembles with the $\mathcal{O}(a)$-improved Wilson fermions~\cite{RQCD:2022xux}. 
The data cover the pion mass from $127$ MeV to $429$ MeV, along three quark-mass trajectories: the SU(3)-symmetric line, a line with fixed sum of the quark masses as a constant, and a line with the strange quark mass kept near its physical value. 
The octet-baryon masses were analyzed using the covariant baryon chiral perturbation theory (BChPT) up to NNLO~\cite{MartinCamalich:2010fp}, where the finite-volume corrections are included.  And, the lattice-spacing effects were parametrized by the leading $a^2$ term and the quark-mass dependent term $a^2M_\phi^2$, where $M_\phi$ denotes the pseudoscalar meson mass. They achieved a rather good description of the lattice data and determined the gradient-flow scale $\sqrt{t_0}=0.1449^{(+0.0007)}_{(-0.0009)}$ fm with high precision.

However, the fitted axial couplings are relatively small with $D+F\approx0.8$, compared with the physical $g_A\approx1.26$, which may hint at sizeable higher-order chiral corrections beyond NNLO.
In addition, the treatment of lattice-spacing effects there deserves further examination. 
Depending on the power counting adopted for the combined expansion in quark mass and lattice spacing, either the $a^2$ or the $a^2M_\phi^2$ terms can first contribute at $\mathcal{O}(p^4)$~\cite{Bar:2003mh}, beyond the $\mathcal{O}(p^3)$ accuracy of the NNLO baryon mass expansion. Taken together, these observations motivate extending the chiral expansion up to $\mathcal{O}(p^4)$, i.e. next-to-next-to-next-to-leading order.

Therefore, in the present work, we carry out a comprehensive analysis of  the RQCD data of octet baryon masses in the SU(3) covariant BChPT at N$^3$LO~\cite{Ren:2012aj}. Finite-volume corrections are included self-consistently, and the discretization effects are considered within the framework of Symanzik effective field theory (EFT)~\cite{Symanzik:1983dc,Symanzik:1983gh}. The fit is performed by converting the baryon masses to the dimensionless quantities via the gradient-flow scale in the chiral limit $\sqrt{8t_{0,\rm ch}}$, which is independent of the quark mass. Then, from the best fit, one can determine the flow scale by reproducing the experimental $\Xi$ mass in the isospin limit. Furthermore, we compare the predicted octet-baryon masses at the physical point with the experimental values, and the nucleon sigma terms, $\sigma_{\pi N}$ and $\sigma_{s N}$, with the direct calculation from three-point correlation functions on the same CLS ensembles \cite{Agadjanov:2023efe}. 

Our paper is organized as follows: In section~\ref{secII}, we present the chiral expression of octet baryon masses and the details of the scale setting method. After that the analysis of RQCD data and the determination of gradient flow scale are given in section~\ref{secIII}, where we also show the predictions of octet baryon masses and sigma terms at the physical point. Finally, we summarize in section~\ref{secIV}.

\section{Theoretical framework}\label{secII}
\subsection{Chiral expression for octet baryon masses}
We briefly summarize the expression of octet baryon masses up to N$^3$LO in covariant BChPT, which can be written symbolically as 
\begin{equation}\label{Eq:mBsymbolic}
	m_B^\chi(M_\phi) = m_0 + m_B^{(2)}(M_\phi) + m_B^{(3)}(M_\phi) + m_B^{(4)}(M_\phi),
\end{equation}
 where $m_0$ is the octet baryon mass in the chiral limit. The contribution $m_B^{(2)}$ collects the tree-level terms linear in the squared pseudoscalar masses  $M_\phi^2$ with three low-energy constants (LECs): $b_0$, $b_D$, and $b_F$.
 Starting from $\mathcal{O}(p^3)$, the loop diagrams contribute to $m_B^{(3)}$  in terms of the axial couplings $D$ and $F$ and the decay constant in the chiral limit $F_0$. 
 At $\mathcal{O}(p^4)$, there are three parts contributing: the tree-level terms proportional to LECs $d_{1,...,5,7,8}$, the tadpole diagrams involving LECs $b_{0,D,F}$ and $b_{1,...,8}$, and the one-loop diagrams with mass insertions. 
 To deal with the power counting breaking terms in the above loop diagrams, we employ the extended-on-mass-shell (EOMS) renormalization scheme~\cite{Gegelia:1999gf,Fuchs:2003qc} which satisfies the analytic structure of the loop amplitudes. The renormalization scale $\mu$ is set to  $\mu=1$ GeV in the following calculation.  
The explicit expressions and the calculation details can be found in Ref.~\cite{Ren:2012aj}.  
Note that the virtual decuplet contributions are not included explicitly. Owing to the large number of LECs at N$^3$LO, their effects can be absorbed into the the octet-only formulation \cite{Ren:2013dzt,Ren:2012aj}.
  
To comparison with LQCD data of octet baryon masses, one needs to take into account the finite volume corrections (FVCs) and the discritization effects. 
Following the treatment in Refs.~\cite{QCDSF-UKQCD:2003hmh,Geng:2011wq}, the FVCs only arise from the loop contributions, giving  
 \begin{equation}
 	m_B^{\chi L^3}(M_\phi, L) = m_0 + m_B^{(2)}(M_\phi) + \left[ m_B^{(3)}(M_\phi) + \Delta m_B^{(3)}(L) \right] +  \left[ m_B^{(4)}(M_\phi) + \Delta m_B^{(4)}(L) \right],
 \end{equation}
  where $\Delta m_B^{(3,4)}(L)$ denote the finite-volume mass shifts, and the corresponding expresses are explicitly given in Refs.~\cite{MartinCamalich:2010fp,Ren:2012aj}. 
 
 In order to describe the discritization effects of LQCD at nonzero lattice spacing $a$, we employ the Symanzik EFT~\cite{Symanzik:1983dc,Symanzik:1983gh} and adopt the power counting~\cite{Bar:2003mh,Tiburzi:2005vy,Ren:2013wxa}
 \begin{equation}
 	a\Lambda_\mathrm{QCD}\sim \frac{m_q}{\Lambda_\mathrm{QCD}} \sim p^2.
 \end{equation}
 For the $\mathcal{O}(a)$-improved fermion action considered here, the discretization effects enter at $\mathcal{O}(p^4)$ are parametrized as 
 \begin{equation}\label{Eq:mBchQCD}
 	m_B^\mathrm{LQCD}(M_\phi, L, a) = m_B^{\chi L^3}(1+c_a a^2),
 \end{equation}
 where $c_a$ parametrizes the leading lattice spacing dependence. 
 
 It is worth mentioning that the chiral expression Eq.~\eqref{Eq:mBchQCD} are the function of the LO pseudoscalar masses $M_\phi \equiv M_{0\phi}$. To maintain a consistent treatment, $M_{0\pi}$ and $M_{0 K}$ are obtained  by perturbatively inverting the NLO meson mass relations. The LO eta mass ($M_{0\eta}$) is then fixed by the Gell-Mann-Okubo relation. The details can be found in the Appendix~\ref{App:LOMphi}.
  
\subsection{Scale setting strategy}   
In this subsection, we outline the main procedure to determine the lattice scale by matching the RQCD data to the chiral expressions, where all quantities are expressed in units of the gradient-flow scale in the chiral limit $\tch$. 
The starting point is the quark-mass dependence of the gradient-flow scale $t_0$ at NLO~\cite{Bar:2013ora},
\begin{equation}
\label{eq:t0NLO}
\frac{t_0}{\tch}
=1+\frac{k_1}{(4\pi F_0)^2}
\bigl(2M_K^2+M_\pi^2\bigr)
+\mathcal O(p^4),
\end{equation}
where $k_1$ is a dimensionless low-energy constant.
We first express the simulated masses in units of the ensemble-dependent scale $\sqrt{8t_{0}}$ and then use Eq.~\eqref{eq:t0expansion} to
convert them to the fixed unit $\sqrt{8\tch}$.
Accordingly, the chiral mass formula can be rescaled in the same fixed unit. Unlike a formulation in the mass-dependent unit $\sqrt{8t_0}$ employed by the RQCD Collaboration~\cite{RQCD:2022xux}, this procedure does not require additional shifts of the LECs, associated with the quark-mass dependence of $t_0$,  in chiral expressions.

\subsubsection{Conversion of the RQCD data}
In Ref.~\cite{RQCD:2022xux}, the RQCD Collaboration reported the lowest-lying octet baryon masses on the CLS $N_f=2+1$ ensembles generated with the 
$\mathcal O(a)$-improved Wilson-clover fermions, and six gauge couplings
$\beta=3.34,\,3.40,\,3.46,\,3.55,\,3.70,\,3.85$. The data set comprises $22$ ensembles on the flavor-symmetric line $m_s=\hat m$ with 
$\hat m=m_u=m_d$, where the octet baryons are exactly degenerate and a single octet mass is determined per ensemble, and $36$ ensembles with $m_s\neq\hat m$, where all four octet masses are determined. The pion and kaon masses, $aM_{\pi,\,K}$, and the gradient-flow quantity
$t_0/a^2$ are listed in Tables~17 and~18 of
Ref.~\cite{RQCD:2022xux}, while the octet baryon masses,
$am_{N,\,\Lambda,\,\Sigma,\,\Xi}$, are listed in Tables~17 and~19.
All of these quantities are given in lattice units.

For the conversion of the lattice data, we first rewrite the above NLO expression of $t_0$ Eq.~\eqref{eq:t0NLO} as
\begin{equation}
\label{eq:t0expansion}
\frac{t_0}{\tch}
=1+\ktt\,\frac{8t_0}{3}
\bigl(2M_K^2+M_\pi^2\bigr)
+\mathcal O(p^4),
\end{equation}
with
\begin{equation*}
\ktt=\frac{3k_1}{8\tch(4\pi F_0)^2},
\end{equation*}
where  $\ktt=-0.0466(62)$ is the continuum slope determined by RQCD~\cite{RQCD:2022xux}, which holds up to $\mathcal{O}(a^2)$. 
In this way, the quark-mass dependence of $t_0$ is expressed through the dimensionless combinations $t_0M_\pi^2$ and
$t_0M_K^2$, which are obtained directly from the lattice simulations  $t_0/a^2$ and $aM_{\pi,K}$. Then, the ratio
$t_0/\tch$ can be determined for each ensemble:
\begin{equation}
\frac{t_{0,i}}{\tch}\simeq1+\ktt X_i,
\qquad
X_i=\frac23\varphi_{4,i},
\qquad
\varphi_{4,i}
=8t_{0,i}\left(M_K^2+\frac12M_\pi^2\right)_i.
\end{equation}
This ratio allows the simulated masses, initially expressed
in units of $\sqrt{8t_0}$, to be converted to the common fixed
unit $\sqrt{8\tch}$, without first determining the absolute lattice scale.
Then, we can obtain the dimensionless octet-baryon masses in terms of $\sqrt{8\tch}$,
\begin{equation}
\tilde m_B^{(i)} =(am_B)^{(i)} \sqrt{8\left(\frac{t_0}{a^2}\right)^{(i)}}
\sqrt{\frac{\tch}{t_{0,i}}}
\simeq (am_B)^{(i)} \sqrt{8\left(\frac{t_0}{a^2}\right)^{(i)}} (1+\ktt X_i)^{-1/2},
\end{equation}
from the raw simulation data in lattice units. 
Since the CLS ensembles are generated at fixed $\beta$, without adjusting the bare coupling to the quark mass, the lattice spacing depends on the quark mass at $\mathcal{O}(a)$~\cite{RQCD:2022xux}. However, our conversion above is free of the $\mathcal{O}(a)$ effects: the masses are first expressed in the ensemble-dependent unit $\sqrt{8t_0}$, in which the lattice spacing cancels, and the remaining factor depends only on the $\mathcal{O}(a)$-free combination $\phi_4$ through the continuum relation Eq.~\eqref{eq:t0expansion}. This is equivalent to the formulation of RQCD, up to higher-order terms in the chiral expansion of $t_0$.

The same conversion applies to the pion and kaon masses,
\begin{equation}
\tilde M_{\pi/K}^{(i)}
\simeq(aM_{\pi/K})^{(i)}\sqrt{8\left(\frac{t_0}{a^2}\right)^{(i)}}(1+\ktt X_i)^{-1/2}.
\end{equation}
Furthermore, the box length is expressed in the same unit as
$\tilde L^{(i)}=L^{(i)}/\sqrt{8\tch}$, with $L^{(i)}=N_s^{(i)}a_i$.

\subsubsection{Dimensionless fit formula}
\label{sec:matchdim}
Next, we prepare the chiral mass formula in the fixed unit
$\sqrt{8\tch}$ to fit the above converted lattice data. One can rescale all dimensional quantities appearing in Eq.~\eqref{Eq:mBchQCD} in the following way, 
\begin{equation}
\begin{aligned}
\tilde m_0&=\sqrt{8\tch}\,m_0, &
\tilde M_{\phi}&=\sqrt{8\tch}\,M_\phi, &
\tilde F_0&=\sqrt{8\tch}\,F_0, \\
\tilde\mu &=\sqrt{8\tch}\,\mu, &
\tilde L&=\frac{L}{\sqrt{8\tch}}.
\end{aligned}
\end{equation}
In addition, all dimensional LECs are transformed accordingly:
\begin{equation}\label{eq:substitution}
	\tilde{b}_{0,D,F,1,...,4} = \frac{b_{0,D,F,1,...,4}}{\sqrt{8\tch}} ,\quad 
	\tilde{b}_{5,...,8} = \frac{b_{5,...,8}}{8\tch},\quad 
	\tilde{d}_{1,...,5,7,8} = \frac{d_{1,...,5,7,8}}{(8\tch)^{3/2}} ,
\end{equation}
 whereas the
dimensionless axial couplings $D$ and $F$ remain unchanged.
In this way, we have the chiral mass formula used to fit the lattice ensemble $i$ 
\begin{equation}
\label{eq:modelfit}
\tilde m_B^{\mathrm{fit}}(i)=
\tilde m_B^{\chi L^3}
\bigl(\tilde M_{0\pi}^{(i)},
\tilde M_{0K}^{(i)},
\tilde M_{0\eta}^{(i)},
\tilde L^{(i)} \bigr) \left(1+\tilde c_a\tilde a_i^2\right),
\end{equation}
which retains the same functional form as Eq.~\eqref{Eq:mBchQCD}. The dependence on the LECs and the renormalization scale is implicit in the above expression. It is worth to noting that the LO meson masses $\tilde M_{0\phi}^{(i)}$ are obtained from the pseudoscalar masses of RQCD as described in the Appendix~\ref{App:LOMphi}.

For the discretization term in Eq.~\eqref{eq:modelfit}, we use the reference flow scale
$\tstar$, defined on the flavor-symmetric line $m_s=\hat m$
by the condition
$\varphi_4^*=12\tstar M_\pi^2=1.11$~\cite{RQCD:2022xux}\footnote{
The ratios $\tstar/a^2$ can be determined accurately by interpolating
between nearby simulation points, providing a convenient reference
for relating different lattice spacings.}. Then, we define 
\begin{equation}
\label{eq:tadef}
\tilde a_i\equiv\frac{a_i}{\sqrt{8\tstar}} =\left[8\left(\frac{\tstar}{a^2}\right)_{\beta_i}\right]^{-1/2},
\qquad
\tilde c_a=8\tstar\,c_a.
\end{equation}
The ratios $t_0^*/a^2(\beta)$ are taken from the global interpolation results of RQCD for different $\beta$:  
\begin{equation}\label{Eq:t0stara2_val}
t_0^* / a^2=2.204, \quad 2.888, \quad 3.686, \quad 5.157, \quad 8.617, \quad 13.988 .
\end{equation}
This normalization allows the discretization term to be evaluated
directly from the dimensionless ratios $\tstar/a^2$. 

\subsubsection{Scale determination}

Finally, in order to set the absolute scale of RQCD simulation, we  choose the $\Xi$ baryon mass at the physical point in the continuum and infinite-volume limits as the input, following Ref.~\cite{RQCD:2022xux}.
As given in FLAG24 review~\cite{FlavourLatticeAveragingGroupFLAG:2024oxs}, the physical point in the quark mass plane is defined by the isospin-symmetric pion and kaon masses  
$M_{\pi,\mathrm{ph}}=134.8(3)$~MeV,
$M_{K,\mathrm{ph}}=494.2(3)$~MeV. The corresponding physical $\Xi$ mass is $m_{\Xi,\mathrm{ph}}=1316.9(3)$~MeV by taking the isospin average and correcting the $\Xi^-$ mass for the electromagnetic effects. 

The chiral-limit flow scale is determined via 
\begin{equation}
\label{eq:calt0ch}
\sqrt{8\tch}=\frac{
\tilde m_\Xi^{\mathrm{fit}}
\bigl(
\tilde M_{0\pi,\mathrm{ph}},
\tilde M_{0K,\mathrm{ph}},
\tilde M_{0\eta,\mathrm{ph}}
\bigr)}
{m_{\Xi,\mathrm{ph}}},
\end{equation}
where the fitted mass is evaluated at $L\to \infty$ and $a\to 0$.
The corresponding LO pion and kaon masses, $\tilde M_{0\pi,\mathrm{ph}}$ and 
$\tilde M_{0K,\mathrm{ph}}$, are obtained from the physical pion and kaon masses in the fixed chiral unit, $\tilde M_{\pi,\mathrm{ph}}$ and $\tilde M_{K,\mathrm{ph}}$, which are defined as 
\begin{equation}
\tilde M_{\pi,\mathrm{ph}}
=\frac{M_{\pi,\mathrm{ph}}}{m_{\Xi,\mathrm{ph}}}
\tilde m_{\Xi,\mathrm{ph}}^{\mathrm{fit}},\quad 
\tilde M_{K,\mathrm{ph}}
=\frac{M_{K,\mathrm{ph}}}{m_{\Xi,\mathrm{ph}}}
\tilde m_{\Xi,\mathrm{ph}}^{\mathrm{fit}},
\end{equation}
by using the inversion formulas in Appendix~\ref{App:LOMphi}.
These coupled relations are solved self-consistently to determine
$\sqrt{8\tch}$.

Once $\tch$ is determined, the gradient flow scale at the physical point $t_{0,\mathrm{ph}}$ is calculated by the NLO parametrization
\begin{equation}\label{eq:calt0ph}
\sqrt{8t_{0,\mathrm{ph}}}
\simeq
\frac{\sqrt{8\tch}}
{\sqrt{1-\frac{\ktt}{3}
\bigl(2\tilde M_{K,\mathrm{ph}}^2
+\tilde M_{\pi,\mathrm{ph}}^2\bigr)}}.
\end{equation}
The reference scale is obtained from
\begin{equation}\label{eq:calt0star}
\sqrt{8\tstar}
\simeq\sqrt{8\tch}\,
\sqrt{1+\frac23\ktt\varphi_4^*}.
\end{equation}
Finally, the lattice spacings are determined from the input ratios
$\tstar/a^2$ through
\begin{equation}
a(\beta)
=\frac{\sqrt{8\tstar}}
{\sqrt{8(\tstar/a^2)_\beta}},
\end{equation}
for different values of $\beta$.

\section{Results and discussion}\label{secIII}
\subsection{Fitting strategy}
As shown in Eq.~(\ref{eq:modelfit}), the chiral expression used to fit the LQCD data has $20$ free parameters to be fixed: 19 scaled LECs $ \tilde m_0$, $\tilde b_0$, $\tilde b_D$, $\tilde b_F$, $
\tilde b_{1,\ldots,8}$, $\tilde d_{1,...,5,7,8}$,
and the discretization coefficient $\tilde c_a$.
The axial couplings are taken as $D=0.80$ and $F=0.46$, which together give  $g_A=D+F=1.26$.
For the pseudoscalar decay constant in the chiral limit, we use $\tilde F_0=\sqrt{8\tch}\,F_0=0.1502$~\cite{Bali:2022qja}.
The renormalization scale is set to $\mu=1$~GeV, with the corresponding
dimensionless scale $\tilde\mu=\sqrt{8\tch}\,\mu$ and the 
LECs in the mesonic sector specified in Appendix~\ref{App:LOMphi}.
The flow-scale inputs $\ktt$ and $\tstar/a^2(\beta)$ are also fixed to the values determined by RQCD as described above.

In order to improve the reliability of the chiral extrapolation, we choose
$\varphi_4=8t_0(M_K^2+\tfrac12M_\pi^2)<1.6$, following the mass cut used in the RQCD analysis~\cite{RQCD:2022xux}, 
to exclude ensembles
with large average quark masses.
To better control the finite-volume effects of RQCD data, we further impose the additional limitation $M_\pi L>3.8$. 
The combined selection retains $43$ ensembles and $127$ baryon masses with $N_{\mathrm{dof}}=107$. For comparison, we also consider two additional data selections, which serve to estimate the systematic uncertainty associated with the finite-volume cut. Using $\varphi_4<1.6$ and $M_\pi L>3.5$ retains $46$ ensembles and $139$ baryon
masses with $N_{\mathrm{dof}}=119$. Imposing the stricter cut $M_\pi L>4.0$ in addition to $\varphi_4<1.6$ retains $41$ ensembles and $122$ baryon masses with $N_{\mathrm{dof}}=102$.

The chi-square function is defined as  
\begin{equation}
\label{eq:chi2}
\chi^2=\sum_i
\bigl[\tilde{\boldsymbol m}^{(i)}
-\tilde{\boldsymbol m}^{\mathrm{fit}}(i)\bigr]^T
\bigl[C^{(i)}\bigr]^{-1}
\bigl[\tilde{\boldsymbol m}^{(i)}
-\tilde{\boldsymbol m}^{\mathrm{fit}}(i)\bigr],
\end{equation}
where $i$ labels the CLS ensembles and
$\tilde{\boldsymbol m}^{(i)}$ contains the converted octet masses
in the fixed chiral unit $\sqrt{8\tch}$. $C^{(i)}$ denotes the covariance matrix for each ensemble, which is $4\times 4$ for ensembles with $m_s\neq\hat m$ and $1\times1$ on the flavor-symmetric line $m_s=\hat m$.
Following the treatment of correlated uncertainties among the octet baryon masses in
Refs.~\cite{MartinCamalich:2010fp,Ren:2012aj}, we have 
\begin{equation}
\label{eq:block}
C^{(i)}_{BB'}=
\bigl(\sigma^{\mathrm{stat}}_{B,i}\bigr)^2\delta_{BB'}
+\bigl(\delta_i^{\mathrm{rel}}\bigr)^2
\tilde m_B^{(i)}\tilde m_{B'}^{(i)},
\qquad
\delta_i^{\mathrm{rel}}\simeq
\frac12\frac{\delta(t_0/a^2)_i}{(t_0/a^2)_i},
\end{equation}
where $\sigma^{\mathrm{stat}}_{B,i}$ is obtained by multiplying the quoted statistical error of $am_B$ by the corresponding mass-conversion factor. The uncertainty of the common flow-scale factor is included separately in the second term.

\begin{table}[t]
\centering
\caption{Values of the dimensionless LECs from the best fit to the RQCD data with the three data selections. The fit $\chi^2$ and $\chi^2/\mathrm{d.o.f.}$ are also given for three fits.}
\label{Tab:FitLECs}
\begin{tabular}{c|ccc}
\hline\hline 
 & Fit-A  & Fit-B&  Fit-C \\
\hline 
$\tilde m_0$ & $1.84152(698)$ & $1.84507(696)$ & $1.88298(647)$ \\
$\tilde b_0$ & $-0.34777(65)$ & $-0.34925(66)$ & $-0.32761(63)$ \\
$\tilde b_D$ & $\phantom{-}0.09363(63)$ & $\phantom{-}0.09331(64)$ & $\phantom{-}0.12663(60)$ \\
$\tilde b_F$ & $-0.17145(106)$ & $-0.16346(107)$ & $-0.19915(99)$ \\
$\tilde b_1$ & $\phantom{-}0.11633(117)$ & $\phantom{-}0.11655(118)$ & $\phantom{-}0.20309(112)$ \\
$\tilde b_2$ & $-0.32664(493)$ & $-0.28743(499)$ & $-0.20597(468)$ \\
$\tilde b_3$ & $-0.52866(170)$ & $-0.56030(172)$ & $-0.49635(164)$ \\
$\tilde b_4$ & $-0.10718(100)$ & $-0.11902(101)$ & $-0.07742(96)$ \\
$\tilde b_5$ & $\phantom{-}0.00094(104)$ & $\phantom{-}0.01150(105)$ & $-0.00218(98)$ \\
$\tilde b_6$ & $\phantom{-}0.06590(439)$ & $\phantom{-}0.06454(443)$ & $-0.17731(408)$ \\
$\tilde b_7$ & $\phantom{-}0.63026(152)$ & $\phantom{-}0.64051(154)$ & $\phantom{-}0.59385(143)$ \\
$\tilde b_8$ & $-0.28008(89)$ & $-0.26771(90)$ & $-0.28737(84)$ \\
$\tilde d_1$ & $\phantom{-}0.00520(49)$ & $\phantom{-}0.00508(49)$ & $\phantom{-}0.00591(46)$ \\
$\tilde d_2$ & $\phantom{-}0.04198(80)$ & $\phantom{-}0.03794(80)$ & $\phantom{-}0.05199(75)$ \\
$\tilde d_3$ & $\phantom{-}0.01049(35)$ & $\phantom{-}0.01154(35)$ & $\phantom{-}0.01537(33)$ \\
$\tilde d_4$ & $\phantom{-}0.02654(138)$ & $\phantom{-}0.02260(138)$ & $\phantom{-}0.02543(130)$ \\
$\tilde d_5$ & $-0.04937(74)$ & $-0.04495(75)$ & $-0.07149(70)$ \\
$\tilde d_7$ & $-0.01315(34)$ & $-0.01320(34)$ & $-0.01154(33)$ \\
$\tilde d_8$ & $-0.02191(53)$ & $-0.02137(53)$ & $-0.01998(50)$ \\
$\tilde{c}_a$ & $-0.67587(3795)$ & $-0.66447(3875)$ & $-0.66300(3578)$ \\
\hline 
$\chi^2$ & $119.9$ & $115.4$ & $142.1$ \\
$\chi^2/\mathrm{d.o.f.}$ & $1.12$ & $1.13$  & $1.19$ \\
\hline\hline 
\end{tabular}

\end{table}

\subsection{Fit results and gradient-flow scale}

In Table~\ref{Tab:FitLECs}, we tabulate the dimensionless LECs and the corresponding $\chi^2/\mathrm{d.o.f.}$ from three best fits to the RQCD octet baryon masses: Fit A with $\varphi_4<1.6$ and $M_\pi L>3.8$, Fit B with $\varphi_4<1.6$ and $M_\pi L>4.0$, and Fit C with $\varphi_4<1.6$ and $M_\pi L>3.5$. One can see that the three fits give a rather good description of RQCD data, with $\chi^2/\mathrm{d.o.f.}\simeq 1.1- 1.2$. 
The resulting dimensionless LECs vary moderately among the three fits. And, the uncertainties of  the fitted LECs are relatively small, reflecting the good quality of the RQCD data. The three quark-mass trajectories of RQCD data provide useful constraints on different combinations of LECs at N$^3$LO.

Then, we can determine the flow scale in the chiral limit $\sqrt{8\tch}$ by using Eq.~\eqref{eq:calt0ch} to reproduce the $\Xi$ mass at the physical point. The scales  $\sqrt{8t_{0,\mathrm{ph}}}$ and $\sqrt{8t_0^*}$ are also obtained via Eqs.~\eqref{eq:calt0ph} and \eqref{eq:calt0star}. Their values from the three fits are listed in Table~\ref{Tab:scales}. We find that the reference scale $\sqrt{8\tstar}$ is close to
$\sqrt{8t_{0,\mathrm{ph}}}$ for all three fits because the reference point $\varphi_4^*=1.11$ lies close to the physical value of $\varphi_4$, as noticed in the RQCD analysis. Although our determined scales are slightly larger than those reported by RQCD~\cite{RQCD:2022xux},  they are consistent with the quoted uncertainties.

\begin{figure}[t]
\centering
\includegraphics[width=0.68\textwidth]{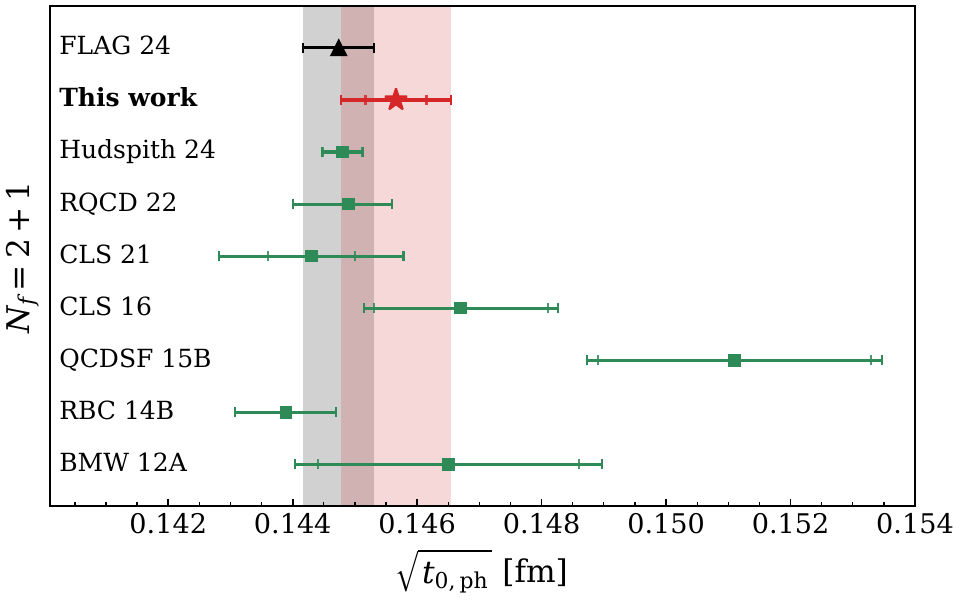}
\caption{Gradient flow scale $\sqrt{t_{0,\mathrm{ph}}}$ from different studies. The light red band is our result, and the gray band denote the averaged value of FLAG24. The others are the $N_f=2+1$ LQCD results  from Hudspith~24~\cite{Hudspith:2024kzk}, RQCD~22~\cite{RQCD:2022xux}, CLS~21~\cite{Strassberger:2021tsu}, CLS~16~\cite{Bruno:2016plf}, QCDSF 15B~\cite{Bornyakov:2015eaa}, RBC~14B~\cite{RBC:2014ntl}, and BMW~12A~\cite{BMW:2012hcm}, repsecitvely.}
\label{Fig:scale}
\end{figure}

\begin{table}[t]
\centering
\caption{Gradient flow scales from three fitting strategies. The corresponding results from RQCD~\cite{RQCD:2022xux} are also listed for comparison.}
\label{Tab:scales}
\setlength{\tabcolsep}{12pt}
\begin{tabular}{c|ccc|c}
\hline\hline 
 & Fit-A & Fit-B & Fit-C & RQCD \\
\hline 
$\sqrt{8\tch}$ [fm]
 & {\bf $0.4192(14)$} & $0.4187(15)$ & $0.4209(13)$ & $0.4170^{+0.0022}_{-0.0027}$ \\
$\sqrt{8t_{0,\mathrm{ph}}}$ [fm]
 & {\bf $0.4120(14)$} & $0.4115(14)$ & $0.4136(13)$ & $0.4098^{+0.0020}_{-0.0025}$ \\
$\sqrt{8\tstar}$ [fm]
 & {\bf $0.4120(14)$} & $0.4115(14)$ & $0.4136(13)$ & $0.4097^{+0.0020}_{-0.0025}$ \\
 \hline\hline 
\end{tabular}
\end{table}
We take $\sqrt{8t_{0,\mathrm{ph}}}$ from Fit-A as the central value and consider the differences among the three fitting strategies as the systematic uncertainty,  
which results in the gradient flow scale at the physical point 
\begin{equation}
\label{eq:scale_result}
\sqrt{8t_{0,\mathrm{ph}}}
=0.4120(14)_{\mathrm{stat.}}(16)_{\mathrm{sym.}}~\mathrm{fm},
\end{equation}
with a total relative uncertainty about $0.5\%$. 
We also present $\sqrt{t_{0,\mathrm{ph}}}$ in Fig.~\ref{Fig:scale} and compare it with the averaged value of FLAG24~\cite{FlavourLatticeAveragingGroupFLAG:2024oxs} and the recent results of $N_f=2+1$ LQCD simulations~\cite{Hudspith:2024kzk,RQCD:2022xux,Strassberger:2021tsu,Bruno:2016plf,Bornyakov:2015eaa,RBC:2014ntl,BMW:2012hcm}. One can see that our result is slightly larger than the FLAG24 average with a difference of $1\sigma$.  

\begin{table}[t]
\centering
\caption{Values of lattice spacings determined in the present work. The lattice spacings from RQCD are also listed for comparison~\cite{RQCD:2022xux}.}
\label{Tab:latticespacings}
\setlength{\tabcolsep}{12pt}
\begin{tabular}{c|c|c}
\hline \hline 
$\beta$ & $a(\beta)$ [fm] & $a^\mathrm{RQCD}(\beta)$ [fm] \\
\hline 
$3.34$ & $0.09810(63)$ & $0.09757(56)$ \\
$3.40$ & $0.08570(55)$ & $0.08524(49)$ \\
$3.46$ & $0.07586(49)$ & $0.07545(44)$ \\
$3.55$ & $0.06413(41)$ & $0.06379(37)$ \\
$3.70$ & $0.04961(32)$ & $0.04934(28)$ \\
$3.85$ & $0.03894(25)$ & $0.03873(22)$ \\
\hline\hline 
\end{tabular}
\end{table}

After that one can obtain the lattice spacings $a$ by using Eq.~\eqref{eq:tadef}, where the ratios  of $t_0^*/a^2$ are taken from the RQCD analysis, as shown in Eq.~\eqref{Eq:t0stara2_val}. 
In Table~\ref{Tab:latticespacings}, we list the lattice spacings at different values of $\beta$. Their uncertainties are obtained by taking into account the statistical and systematical uncertainties of both $\sqrt{8\tch}$ and $t_0^*/a^2$. Our determined lattice spacings are about $1\%$ larger than those obtained by RQCD.

\subsection{Chiral, continuum, and infinite-volume extrapolations }
Next, we are in the place to discussion the chiral, continuum, and infinite-volume extrapolations of the RQCD data with our N$^3$LO chiral expression. 
First, we show the fitted LECs, which are converted to the physical units according to Eq.~\eqref{eq:substitution}, in Table~\ref{tab:lecs}. Here, we only use the $\sqrt{8\tch}$ from Fit-A to perform the conversion. 
Accordingly, we have the pion decay constant in the chiral limit $F_0\simeq 70.7$ MeV and the renormalization scale $\mu \simeq 1.0$~GeV. Note that the octet baryon mass in the chiral limit is
$m_0=0.8667(37)$~GeV, which is close to the values determined in our previous studies~\cite{Ren:2012aj,Ren:2014vea,Ren:2016aeo}. Such value is also numerically close to  the one determined in the SU(2) calculation of nucleon mass at N$^3$LO~\cite{Ren:2017fbv} and beyond~\cite{Chen:2024twu,Liang:2025cjd}  within the covariant BChPT.

\begin{table}[t]
\centering
\caption{LECs from Fit-A in the physical units with the full-covariance statistical errors.}
\label{tab:lecs}
\setlength{\tabcolsep}{12pt}
\begin{tabular}{cc|cc}
\hline\hline
LECs & Fit-A & LECs & Fit-A \\
\hline
$m_0$~[GeV] & $0.8667(37)$ 
 & $b_7$~[GeV$^{-2}$] & $\phantom{-}2.8454(207)$ \\
\cline{1-2} 
$b_0$~[GeV$^{-1}$] & $-0.7389(29)$
 & $b_8$~[GeV$^{-2}$] & $-1.2645(92)$ \\
  \cline{3-4} 
$b_D$~[GeV$^{-1}$] & $\phantom{-}0.1989(15)$
 & $d_1$~[GeV$^{-3}$] & $\phantom{-}0.0499(46)$ \\
$b_F$~[GeV$^{-1}$] & $-0.3643(26)$
 & $d_2$~[GeV$^{-3}$] & $\phantom{-}0.4027(78)$ \\
 \cline{1-2} 
$b_1$~[GeV$^{-1}$] & $\phantom{-}0.2472(25)$
 & $d_3$~[GeV$^{-3}$] & $\phantom{-}0.1006(35)$ \\
$b_2$~[GeV$^{-1}$] & $-0.6940(106)$
 & $d_4$~[GeV$^{-3}$] & $\phantom{-}0.2546(132)$ \\
$b_3$~[GeV$^{-1}$] & $-1.1233(50)$
 & $d_5$~[GeV$^{-3}$] & $-0.4736(84)$ \\
$b_4$~[GeV$^{-1}$] & $-0.2277(22)$
 & $d_7$~[GeV$^{-3}$] & $-0.1261(36)$ \\
$b_5$~[GeV$^{-2}$] & $\phantom{-}0.0042(47)$
 & $d_8$~[GeV$^{-3}$] & $-0.2102(57)$ \\
  \cline{3-4} 
$b_6$~[GeV$^{-2}$] & $\phantom{-}0.2975(197)$
 & $c_a$~[GeV$^2$] & $-0.1550(88)$ \\
\hline\hline
\end{tabular}
\end{table}

\begin{figure}[t]
\centering
\includegraphics[width=0.8\textwidth]{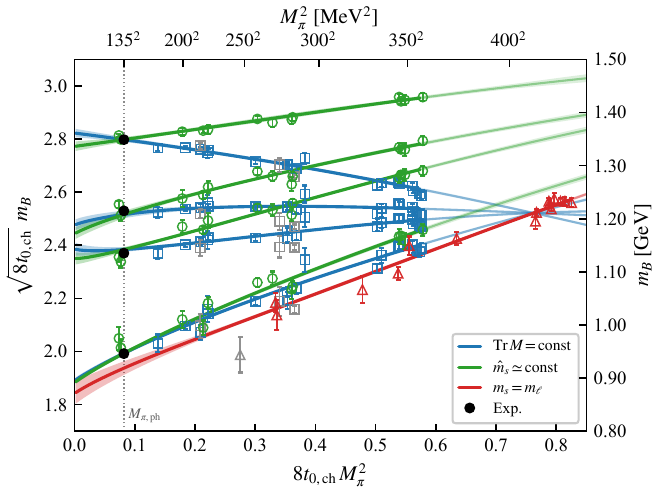}
\caption{Octet baryon masses as functions of the pion mass with the three quark-mass trajectories as given in RQCD. From bottom to top, the curves  are for $N$, $\Lambda$, $\Sigma$ and $\Xi$, and present the continuum and infinite-volume results with the LECs fixed from Fit-A.
The lattice data are corrected to consider the finite-volume correction and the discretization effects, and shifted to the reference trajectories: the trace of the bare quark-mass matrix fixed $\mathrm{Tr}(M) = \mathrm{const}$ (blue squares), the strange-quark mass fixed near
its physical value (green circles), and the flavor-symmetric line $m_s=\hat m$ (red triangles).  The lattice data with $M_\pi L\leq3.8$, excluded from the Fit-A, are denoted by gray points.  Black circles indicate the experimental octet baryon masses by taking the isospin-symmetric corrections, as given in the text.}
\label{Fig:chiraltrajectories}
\end{figure}

In Fig.~\ref{Fig:chiraltrajectories}, we present the octet baryon masses as functions of the squared pion mass ($8\tch\, M_\pi^2$  vs. $\sqrt{8\tch}m_B$ on the bottom and left axes, $M_\pi^2$ vs. $m_B$ on the top and right axes) using the LECs from Fit-A. The strange-quark masses are set according to the quark mass trajectories as shown in Fig.~\ref{Fig:chiraltrajectories}.  
One can see that our N$^3$LO results give a  good description of the lattice data, particularly for the data (with $\varphi_4>1.6$) excluded from the fit. Moreover, the linear dependence of the lattice data on $M_\pi^2$ at large light quark masses is clearly seen in the present study.  
After performing the chiral extrapolation to the physical point, we can predict the octet baryon masses:
$m_N=937.9(35)(28)$ MeV, $m_\Lambda=1121.9(28)(3)$ MeV, $m_\Sigma=1184.1(26)(33)$ MeV. The nucleon mass is consistent with the 
experimental value $m_N=937.54(6)$ MeV in the isospin limit, and the $\Lambda$ ($\Sigma$) mass is slightly larger (smaller) than the isospin-limit value $m_\Lambda=1115.68(1)$ MeV ($m_\Sigma=1190.66(12)$)~\cite{RQCD:2022xux}. 

\begin{figure}[t]
\centering
\includegraphics[width=0.49\textwidth]{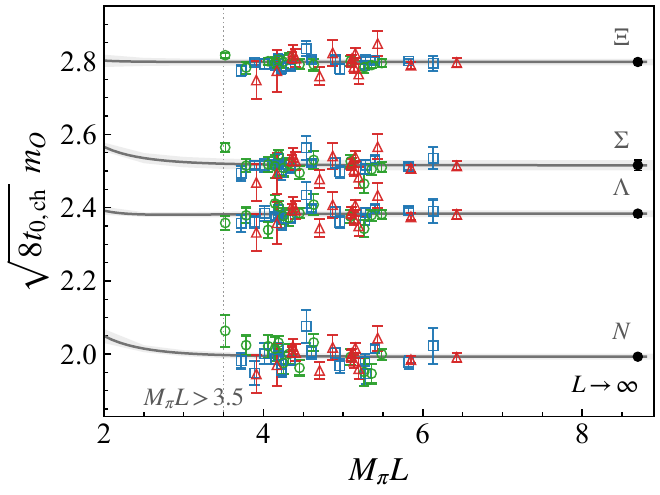}
\includegraphics[width=0.49\textwidth]{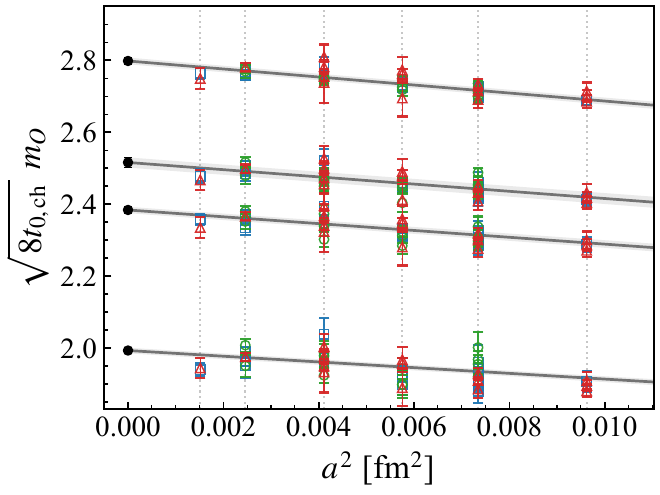}
\caption{Left panel: Octet baryon masses as functions of $M_\pi L$. Right panel: Octet baryon masses as functions of the squared lattice spacing $a^2$. Theoretical curves represent the results of Fit-A in terms of the physical values of pion and kaon masses. Lattice data notation is the same as the one given in Fig.~\ref{Fig:chiraltrajectories}.}
\label{Fig:volume_continuum}
\end{figure}

We also show the volume dependence (left panel) and the lattice spacing dependence (right panel) in Fig.~\ref{Fig:volume_continuum}, where the lattice data are projected to the physical point in terms of the pion and kaon masses. Furthermore, the data are extrapolated  to the continuum limit in the left subfigure, and extrapolated to the infinite volume limit in the right subfigure. 
Basically, the finite volume dependence is mild with $M_\pi L >3$. But including these effects is necessary to improve the description of lattice data and essential to better constrain the values of LECs. 
The discretization effects are sizeable and show an approximately linear dependence on $a^2$, as described by the single  $c_a a^2$ term in  Eq.~\eqref{eq:modelfit}. A similar behavior was also observed in the RQCD analysis  where additional quark-mass-dependent $a^2$ terms were included~\cite{RQCD:2022xux}. To examine their impact, we also include an $a^2\overline M^2$ term with $\overline M^2=(2M_{K}^2+M_\pi^2)/3$, which formally contributes at $\mathcal O(p^6)$ in our power counting, and find only a slight improvement in $\chi^2$. The resulting baryon masses and gradient-flow scale remain almost unchanged, suggesting that the leading $a^2$ term provides an adequate description of the discretization effects at the current order.

\subsection{Baryon sigma terms}
\label{sec:sigma}
Finally, we calculate the sigma terms of octet baryons by using the Feynman-Hellmann theorem with the chiral expression of the octet baryon masses,
\begin{equation}
\label{eq:sigma}
\sigma_{\pi B} \;=\; \hat m\,\frac{\partial m_B}{\partial \hat m},
\qquad
\sigma_{s B} \;=\; m_s\,\frac{\partial m_B}{\partial m_s},
\end{equation}
where $\hat{m}=m_u=m_d$. Using the NLO ChPT, the quark masses can be expressed as functions of $M_{4\pi}$ and $M_{4K}$, as shown in Appendix~\ref{App:LOMphi}. Then, one needs to consider the derivative chain $\partial m_B/\partial m_q = (\partial M^2_{4\pi,K}/\partial m_q) (\partial m_B/\partial M^2_{4\pi,K}$) to obtain the baryon sigma terms. The results are listed in Table~\ref{tab:sigma} using the LECs values from Fit-A. The systematic uncertainties are estimated from the differences among the three fits. 

\begin{table}[t]
\centering
\caption{Predicted pion- and strangeness-sigma terms of the octet baryons (in units of MeV) by the N$^3$LO BChPT with the LECs from Fit-A, and compared with the RQCD results~\cite{RQCD:2022xux}.}
\label{tab:sigma}
\setlength{\tabcolsep}{12pt}
\begin{tabular}{l|cccc} 
\hline \hline 
 & \multicolumn{2}{c}{$\sigma_{\pi B}$} & \multicolumn{2}{c}{$\sigma_{sB}$} \\
\cline{2-3}\cline{4-5}
  & Fit-A & RQCD & Fit-A & RQCD \\
\hline 
$N$ & $45(1)(1)$ & $43.9^{(4.7)}_{(4.7)}$ & $49(10)(6)$ & $16^{(68)}_{(58)}$ \\
$\Lambda$ & $21(1)(4)$ & $28.2^{(5.4)}_{(4.3)}$ & $218(7)(3)$ & $144^{(76)}_{(58)}$ \\
$\Sigma$ & $28(1)(4)$ & $25.9^{(6.1)}_{(3.8)}$ & $233(7)(11)$ & $229^{(70)}_{(65)}$ \\
$\Xi$ & $12(1)(4)$ & $11.2^{(6.4)}_{(4.5)}$ & $341(7)(19)$ & $311^{(83)}_{(72)}$ \\
\hline\hline 
\end{tabular}
\end{table}

The pion-nucleon sigma term, $\sigma_{\pi N}=45(1)(1)$ is in reasonable agreement with the RQCD result $43.9\pm4.7$~MeV~\cite{RQCD:2022xux}
and the direct determination $43.7(3.6)$~MeV by computing the three-point correlation functions on the CLS ensembles~\cite{Agadjanov:2023efe}. Note that this result is smaller than that extracted from the pion-nucleon scattering data, $\sigma_{\pi N}=59.1(3.5)$ MeV~\cite{Hoferichter:2015hva}.
Our predicted strangeness-nucleon sigma term, $\sigma_{sN}=49(10)(6)$~MeV is larger than the central value of the RQCD result and the direct determination of~$28.6(9.3)$~MeV, but remains compatible with both within the quoted uncertainties\footnote{It is worth noting that the N$^3$LO chiral analysis of RQCD data with the large $N_c$ constraint gives a large negative value $\sigma_{sN}=-316 \pm 76$ MeV~\cite{Lutz:2023xpi}.}. For the hyperons, the pion- and strangeness-sigma terms are consistent with the RQCD results. 

\begin{figure}[t]
\centering
\includegraphics[width=0.99\textwidth]{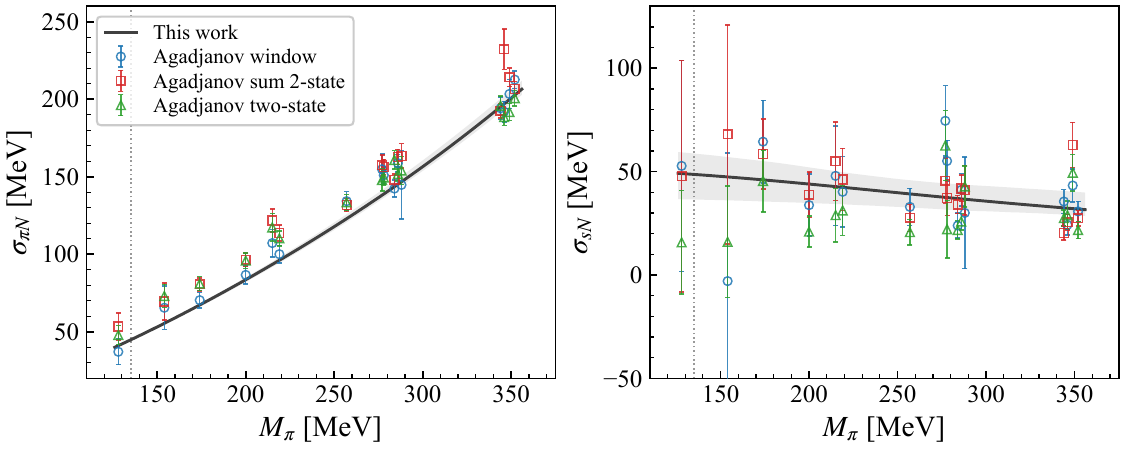}
\caption{Pion-nucleon (left) and strangeness-nucleon (right) sigma terms as
functions of $M_\pi$. The black curves are the predictions from Fit-A along the $\mathrm{Tr}M=\mathrm{const}$ quark mass trajectory in the continuum and infinite-volume limits. 
The data points show the direct evaluation results of Ref.~\cite{Agadjanov:2023efe}, which are shifted by including the finite-volume corrections and the lattice spacing effects.}
\label{fig:sigmampi}
\end{figure}

Furthermore, in Fig.~\ref{fig:sigmampi} we show the pion mass dependence of $\sigma_{\pi N}$ and $\sigma_{sN}$ from the N$^3$LO best fit and compare them with the direct determinations of the sigma terms on 16 CLS ensembles~\cite{Agadjanov:2023efe}. One can see that the agreement is very good. The direct results, particularly the window results, broadly follow the predicted pion-mass dependence within their uncertainties. 

\section{Summary and Outlook}\label{secIV}

We have performed the scale setting of 
the CLS $N_f=2+1$ ensembles by analyzing the RQCD lattice data of octet baryon masses within the EOMS BChPT up to N$^3$LO. Finite-volume corrections are included self-consistently, and the leading $a^2$ discretization effects are considered at the same chiral order. The fit is performed by converting the octet baryon masses into dimensionless quantities through the gradient-flow scale in the chiral limit $\sqrt{8\tch}$. From the best fit, one can estimate the flow scale in the physical point via the experimental $\Xi$ mass in the isospin limit. Finally, we have the gradient flow scale $\sqrt{8t_{0,\mathrm{ph}}}=0.4120(14)(16)$ fm, corresponding to a total precision of $0.5\%$. The resulting lattice spacings are slightly larger than those obtained by the RQCD Collaboration. Through the chiral extrapolation to the physical point, we predict the octet-baryon masses, which are agree well with the experimental values in the isospin limit. 
Furthermore, we calculate the pion- and strangeness-sigma terms of octet baryons via the Feynman-Hellmann theorem. The predicted nucleon sigma terms, $\sigma_{\pi N}$ and $\sigma_{sN}$, are consistent with the results of the direct evaluation via the three-point correlation functions on the same CLS ensembles.

 As a natural extension, we are working on the complementary determination of the gradient-flow scale $t_0$ using the high-precision $\Omega$-baryon mass data of Hudspith {\it et al.}~\cite{Hudspith:2024kzk}, based on the N$^3$LO covariant BChPT in the EOMS scheme for decuplet baryon masses~\cite{Ren:2013oaa}.
Besides, the present framework can be further employed to determine the gradient-flow scale on other $N_f=2+1$ lattice ensembles, such as~\cite{Hu:2024mas,Yoo:2026uul}.

\acknowledgements 
We thank Gunnar S. Bali and Sara Collins for valuable discussions, and ECT$^*$ for its hospitality during the workshop ``Scale Setting: Precision Lattice QCD for Particle and Nuclear Physics''. 
This work has been supported in part by the Key R\&D Program of Shandong Province (Grant No. 2026CXPT175), by the Shandong Provincial Natural Science Fund for Excellent Young Scientists Fund Program (Overseas) with project no. 2025HWYQ-015, and by Qilu Youth Scholars Program of Shandong University.

\appendix

\section{Conversion of pseudoscalar masses and meson inputs}
\label{App:LOMphi}

In this appendix, we calculate the LO pseudoscalar meson masses $M_{0\phi}$ used in the octet-baryon mass expressions by using the corresponding masses obtained from lattice simulations $M_\phi$. 

First, we convert the pion and kaon masses to the dimensionless quantities, $\tilde M_\pi=\sqrt{8\tch}M_\pi \equiv \tilde{M}_{4\pi}$ and $\tilde M_K=\sqrt{8\tch}M_K\equiv \tilde{M}_{4K}$ by using the gradient-flow scale $\sqrt{8\tch}$ in the chiral limit. The subscript $4$ indicates the lattice data is taken as the one with the chiral order $\mathcal O(p^4)$.
Then, one can obtain the LO masses $\tilde M_{0\pi}$ and $\tilde M_{0K}$ by perturbatively inverting the NLO meson mass relations~\cite{Gasser:1984gg,RQCD:2022xux},
\begin{equation}
\label{eq:gmorpi}
\tilde M_{0\pi}^2=\tilde M_{4\pi}^2\Bigl\{1-\tilde\mu_\pi+\tfrac13\tilde\mu_\eta
-\tfrac{8}{\tilde F_0^2}\bigl[\tilde M_{4\pi}^2L_{85}
+(2\tilde M_{4K}^2+\tilde M_{4\pi}^2)L_{64}\bigr]\Bigr\},
\end{equation}
and
\begin{equation}
\label{eq:gmork}
\tilde M_{0K}^2=\tilde M_{4K}^2\Bigl\{1-\tfrac23\tilde\mu_\eta
-\tfrac{8}{\tilde F_0^2}\bigl[\tilde M_{4K}^2L_{85}
+(2\tilde M_{4K}^2+\tilde M_{4\pi}^2)L_{64}\bigr]\Bigr\},
\end{equation}
where chiral logarithms are defined as
\begin{equation*}
\tilde\mu_P=
\frac{\tilde M_{4\phi}^2}{32\pi^2\tilde F_0^2}
\ln\frac{\tilde M_{4\phi}^2}{\tilde\mu^2},
\qquad \phi=\pi,\eta,
\end{equation*}
with the eta mass $\tilde M_{4\eta}^2=(4\tilde M_{4K}^2-\tilde M_{4\pi}^2)/3$, the meson decay constant in the chiral limit $\tilde F_0=\sqrt{8\tch}F_0$, and the renormalization scale $\tilde\mu=\sqrt{8\tch}\mu$.
In the above expression, the LECs combinations are defined as $L_{85}=2L_8-L_5$ and $L_{64}=2L_6-L_4$. After obtain $\tilde M_{0\pi}$ and $\tilde M_{0K}$, one can have the LO eta mass entering the baryon mass expressions 
\begin{equation}
\tilde M_{0\eta}^2=
\frac{4\tilde M_{0K}^2-\tilde M_{0\pi}^2}{3}.
\end{equation}

For the meson LECs, we adopt the ratios determined in Ref.~\cite{RQCD:2022xux}, Eq.~(C.26), at $\mu=770$~MeV,
\begin{equation}
\label{eq}
\frac{L_{85}}{F_0^2}=7.8(4.8)\times10^{-9},\mathrm{MeV}^{-2},
\qquad
\frac{L_{64}}{F_0^2}=-1.5(3.2)\times10^{-9},\mathrm{MeV}^{-2}.
\end{equation}
To determine $L_{85}$ and $L_{64}$, we use $\tilde F_0=0.1502$~\cite{Bali:2022qja} and the reference value $\sqrt{8\tch}=0.4170$~fm~\cite{RQCD:2022xux}, which gives $F_0\simeq71.1$~MeV. This reference value is used only to determine the external meson LECs. 
These combinations are then evolved to the renormalization scale $\mu=1$~GeV according to
\begin{equation}
\label{eq}
L_{85}(\mu')=L_{85}(\mu)-\frac{1}{12}\frac{\ln(\mu^2/\mu'^2)}{16\pi^2},
\qquad
L_{64}(\mu')=L_{64}(\mu)+\frac{1}{72}\frac{\ln(\mu^2/\mu'^2)}{16\pi^2},
\end{equation}
yielding $L_{85}=0.315(24)\times10^{-3}$ and $L_{64}=-0.054(16)\times10^{-3}$. The dimensionless renormalization scale is $\tilde\mu=\sqrt{8\tch}\mu$, which gives $\tilde\mu\simeq2.113$ at $\mu=1$~GeV when evaluated using the reference value of $\sqrt{8\tch}$ above. 

\bibliographystyle{apsrev4-1}
\bibliography{scalerefs}

@article{Ren:2013dzt,
	archiveprefix = {arXiv},
	author = {Ren, Xiu-Lei and Geng, Lisheng and Meng, Jie and Toki, Hiroshi},
	doi = {10.1103/PhysRevD.87.074001},
	eprint = {1302.1953},
	journal = {Phys. Rev. D},
	number = {7},
	pages = {074001},
	primaryclass = {nucl-th},
	title = {{Virtual decuplet effects on octet baryon masses in covariant baryon chiral perturbation theory}},
	volume = {87},
	year = {2013}}

@article{Ren:2017fbv,
	archiveprefix = {arXiv},
	author = {Ren, Xiu-Lei and Ling, Xi-Zhe and Geng, Li-Sheng},
	doi = {10.1016/j.physletb.2018.05.063},
	eprint = {1710.07164},
	journal = {Phys. Lett. B},
	pages = {7--12},
	primaryclass = {hep-ph},
	title = {{Pion{\textendash}nucleon sigma term revisited in covariant baryon chiral perturbation theory}},
	volume = {783},
	year = {2018}}

@article{Strassberger:2021tsu,
	archiveprefix = {arXiv},
	author = {Strassberger, Ben and others},
	doi = {10.22323/1.396.0135},
	eprint = {2112.06696},
	journal = {PoS},
	pages = {135},
	primaryclass = {hep-lat},
	reportnumber = {DESY-21-175, WUB/21-05},
	title = {{Scale setting for CLS 2+1 simulations}},
	volume = {LATTICE2021},
	year = {2022}}

@article{Bornyakov:2015eaa,
	archiveprefix = {arXiv},
	author = {Bornyakov, V. G. and others},
	eprint = {1508.05916},
	month = {8},
	primaryclass = {hep-lat},
	reportnumber = {ADP-15-25-T927, DESY-15-154, EDINBURGH-2015-19, LIVERPOOL-LTH-1054},
	title = {{Wilson flow and scale setting from lattice QCD}},
	year = {2015}}

@article{Lutz:2023xpi,
	archiveprefix = {arXiv},
	author = {Lutz, Matthias F. M. and Heo, Yonggoo and Guo, Xiao-Yu},
	doi = {10.1140/epjc/s10052-023-11556-1},
	eprint = {2301.06837},
	journal = {Eur. Phys. J. C},
	number = {5},
	pages = {440},
	primaryclass = {hep-lat},
	title = {{Low-energy constants in the chiral Lagrangian with baryon octet and decuplet fields from Lattice QCD data on CLS ensembles}},
	volume = {83},
	year = {2023}}

@article{Liang:2025cjd,
	archiveprefix = {arXiv},
	author = {Liang, Ze-Rui and Chen, Han-Xue and Guo, Feng-Kun and Guo, Zhi-Hui and Yao, De-Liang},
	doi = {10.1007/JHEP04(2025)192},
	eprint = {2502.19168},
	journal = {JHEP},
	pages = {192},
	primaryclass = {hep-ph},
	title = {{Chiral representation of the nucleon mass at leading two-loop order}},
	volume = {04},
	year = {2025}}

@article{Chen:2024twu,
	archiveprefix = {arXiv},
	author = {Chen, Long-Bin and Hu, Siwei and Jia, Yu and Mo, Zhewen},
	eprint = {2406.04124},
	month = {6},
	primaryclass = {hep-ph},
	title = {{Light quark mass dependence of nucleon mass to two-loop order}},
	year = {2024}}

@article{Ren:2016aeo,
	archiveprefix = {arXiv},
	author = {Ren, Xiu-Lei and Alvarez-Ruso, L. and Geng, Li-Sheng and Ledwig, T. and Meng, Jie and Vicente Vacas, M. J.},
	doi = {10.1016/j.physletb.2017.01.024},
	eprint = {1606.03820},
	journal = {Phys. Lett. B},
	pages = {325--333},
	primaryclass = {nucl-th},
	title = {{Consistency between SU(3) and SU(2) covariant baryon chiral perturbation theory for the nucleon mass}},
	volume = {766},
	year = {2017}}

@article{Ren:2014vea,
	archiveprefix = {arXiv},
	author = {Ren, Xiu-Lei and Geng, Li-Sheng and Meng, Jie},
	doi = {10.1103/PhysRevD.91.051502},
	eprint = {1404.4799},
	journal = {Phys. Rev. D},
	number = {5},
	pages = {051502},
	primaryclass = {hep-ph},
	title = {{Scalar strangeness content of the nucleon and baryon sigma terms}},
	volume = {91},
	year = {2015}}

@article{Hudspith:2024kzk,
	archiveprefix = {arXiv},
	author = {Hudspith, Renwick J. and Lutz, Matthias F. M. and Mohler, Daniel},
	eprint = {2404.02769},
	month = {4},
	primaryclass = {hep-lat},
	title = {{Precise Omega baryons from lattice QCD}},
	year = {2024}}

@article{Yoo:2026uul,
	archiveprefix = {arXiv},
	author = {Yoo, Jun-sik and Ee, June-Haak and Park, Sungwoo and Gupta, Rajan and Bhattacharya, Tanmoy and Mondal, Santanu and Jo{\'o}, B{\'a}lint and Edwards, Robert and Orginos, Kostas and Winter, Frank},
	eprint = {2601.10857},
	month = {1},
	primaryclass = {hep-lat},
	reportnumber = {LA-UR-25-32114},
	title = {{The Spectrum and Scale Setting on 2+1-flavor NME Lattices}},
	year = {2026}}

@article{Hu:2024mas,
	archiveprefix = {arXiv},
	author = {Hu, Bolun and Du, Haiyang and Jiang, Xiangyu and Liu, Keh-Fei and Sun, Peng and Yang, Yi-Bo},
	eprint = {2411.18402},
	month = {11},
	primaryclass = {hep-lat},
	title = {{Unveiling the Strong Interaction origin of Baryon Masses with Lattice QCD}},
	year = {2024}}

@article{Ren:2013oaa,
	archiveprefix = {arXiv},
	author = {Ren, Xiu-Lei and Geng, Li-Sheng and Meng, Jie},
	doi = {10.1103/PhysRevD.89.054034},
	eprint = {1307.1896},
	journal = {Phys. Rev. D},
	number = {5},
	pages = {054034},
	primaryclass = {nucl-th},
	title = {{Decuplet baryon masses in covariant baryon chiral perturbation theory}},
	volume = {89},
	year = {2014}}

@article{Hoferichter:2015hva,
	archiveprefix = {arXiv},
	author = {Hoferichter, Martin and Ruiz de Elvira, Jacobo and Kubis, Bastian and Mei{\ss}ner, Ulf-G.},
	doi = {10.1016/j.physrep.2016.02.002},
	eprint = {1510.06039},
	journal = {Phys. Rept.},
	pages = {1--88},
	primaryclass = {hep-ph},
	reportnumber = {INT-PUB-15-050},
	title = {{Roy{\textendash}Steiner-equation analysis of pion{\textendash}nucleon scattering}},
	volume = {625},
	year = {2016}}

@article{Bali:2022qja,
	archiveprefix = {arXiv},
	author = {Bali, Gunnar S. and Collins, Sara and S{\"o}ldner, Wolfgang and Weish{\"a}upl, Simon},
	collaboration = {RQCD},
	doi = {10.1103/PhysRevD.105.054516},
	eprint = {2201.05591},
	journal = {Phys. Rev. D},
	number = {5},
	pages = {054516},
	primaryclass = {hep-lat},
	title = {{Leading order mesonic and baryonic SU(3) low energy constants from Nf=3 lattice QCD}},
	volume = {105},
	year = {2022}}

@article{Gasser:1984gg,
	author = {Gasser, J. and Leutwyler, H.},
	doi = {10.1016/0550-3213(85)90492-4},
	journal = {Nucl. Phys. B},
	pages = {465--516},
	reportnumber = {CERN-TH-3798},
	title = {{Chiral Perturbation Theory: Expansions in the Mass of the Strange Quark}},
	volume = {250},
	year = {1985}}

@book{Rothe:1992nt,
	author = {Rothe, Heinz J.},
	doi = {10.1142/8229},
	isbn = {978-981-4365-87-1, 978-981-4365-85-7},
	publisher = {World Scientific Publishing Company},
	title = {{Lattice Gauge Theories : An Introduction (Fourth Edition)}},
	volume = {43},
	year = {2012}}

@book{Gattringer:2010zz,
	address = {Berlin},
	author = {Gattringer, Christof and Lang, Christian B.},
	doi = {10.1007/978-3-642-01850-3},
	isbn = {978-3-642-01849-7, 978-3-642-01850-3},
	publisher = {Springer},
	title = {{Quantum chromodynamics on the lattice}},
	volume = {788},
	year = {2010}}

@article{FlavourLatticeAveragingGroupFLAG:2024oxs,
	archiveprefix = {arXiv},
	author = {Aoki, Y. and others},
	collaboration = {Flavour Lattice Averaging Group (FLAG)},
	doi = {10.1103/nfzp-p5dn},
	eprint = {2411.04268},
	journal = {Phys. Rev. D},
	number = {1},
	pages = {014508},
	primaryclass = {hep-lat},
	reportnumber = {CERN-TH-2024-192, FERMILAB-PUB-24-0785-T},
	title = {{FLAG review 2024}},
	volume = {113},
	year = {2026}}

@article{Sommer:1993ce,
	archiveprefix = {arXiv},
	author = {Sommer, R.},
	doi = {10.1016/0550-3213(94)90473-1},
	eprint = {hep-lat/9310022},
	journal = {Nucl. Phys. B},
	pages = {839--854},
	reportnumber = {DESY-93-062},
	title = {{A New way to set the energy scale in lattice gauge theories and its applications to the static force and $\alpha_s$ in SU(2) Yang-Mills theory}},
	volume = {411},
	year = {1994}}

@article{Bernard:2000gd,
	archiveprefix = {arXiv},
	author = {Bernard, Claude W. and Burch, Tom and Orginos, Kostas and Toussaint, Doug and DeGrand, Thomas A. and DeTar, Carleton E. and Gottlieb, Steven A. and Heller, Urs M. and Hetrick, James E. and Sugar, Bob},
	doi = {10.1103/PhysRevD.62.034503},
	eprint = {hep-lat/0002028},
	journal = {Phys. Rev. D},
	pages = {034503},
	title = {{The Static quark potential in three flavor QCD}},
	volume = {62},
	year = {2000}}

@article{Luscher:2010iy,
	archiveprefix = {arXiv},
	author = {L{\"u}scher, Martin},
	doi = {10.1007/JHEP08(2010)071},
	eprint = {1006.4518},
	journal = {JHEP},
	note = {[Erratum: JHEP 03, 092 (2014)]},
	pages = {071},
	primaryclass = {hep-lat},
	reportnumber = {CERN-PH-TH-2010-143},
	title = {{Properties and uses of the Wilson flow in lattice QCD}},
	volume = {08},
	year = {2010}}

@article{BMW:2012hcm,
	archiveprefix = {arXiv},
	author = {Bors{\'a}nyi, Szabolcs and D{\"u}rr, Stephan and Fodor, Zolt{\'a}n and Hoelbling, Christian and Katz, S{\'a}ndor D. and Krieg, Stefan and Kurth, Thorsten and Lellouch, Laurent and Lippert, Thomas and McNeile, Craig},
	collaboration = {BMW},
	doi = {10.1007/JHEP09(2012)010},
	eprint = {1203.4469},
	journal = {JHEP},
	pages = {010},
	primaryclass = {hep-lat},
	reportnumber = {ITP-BUDAPEST-657, CPT-P004-2012, WUB-12-02},
	title = {{High-precision scale setting in lattice QCD}},
	volume = {09},
	year = {2012}}

@article{Bar:2013ora,
	archiveprefix = {arXiv},
	author = {Bar, Oliver and Golterman, Maarten},
	doi = {10.1103/PhysRevD.89.034505},
	eprint = {1312.4999},
	journal = {Phys. Rev. D},
	note = {[Erratum: Phys.Rev.D 89, 099905 (2014)]},
	number = {3},
	pages = {034505},
	primaryclass = {hep-lat},
	title = {{Chiral perturbation theory for gradient flow observables}},
	volume = {89},
	year = {2014}}

@article{RQCD:2022xux,
	archiveprefix = {arXiv},
	author = {Bali, Gunnar S. and Collins, Sara and Georg, Peter and Jenkins, Daniel and Korcyl, Piotr and Sch{\"a}fer, Andreas and Scholz, Enno E. and Simeth, Jakob and S{\"o}ldner, Wolfgang and Weish{\"a}upl, Simon},
	collaboration = {RQCD},
	doi = {10.1007/JHEP05(2023)035},
	eprint = {2211.03744},
	journal = {JHEP},
	pages = {035},
	primaryclass = {hep-lat},
	title = {{Scale setting and the light baryon spectrum in N$_{f}$ = 2 + 1 QCD with Wilson fermions}},
	volume = {05},
	year = {2023}}

@article{Bruno:2016plf,
	archiveprefix = {arXiv},
	author = {Bruno, Mattia and Korzec, Tomasz and Schaefer, Stefan},
	doi = {10.1103/PhysRevD.95.074504},
	eprint = {1608.08900},
	journal = {Phys. Rev. D},
	number = {7},
	pages = {074504},
	primaryclass = {hep-lat},
	reportnumber = {DESY-16-162, WUB-16-05},
	title = {{Setting the scale for the CLS $2 + 1$ flavor ensembles}},
	volume = {95},
	year = {2017}}

@article{RBC:2014ntl,
	archiveprefix = {arXiv},
	author = {Blum, T. and others},
	collaboration = {RBC, UKQCD},
	doi = {10.1103/PhysRevD.93.074505},
	eprint = {1411.7017},
	journal = {Phys. Rev. D},
	number = {7},
	pages = {074505},
	primaryclass = {hep-lat},
	reportnumber = {KEK-TH-1769, RBRC-1095, DAMTP-2014-86},
	title = {{Domain wall QCD with physical quark masses}},
	volume = {93},
	year = {2016}}

@article{FlavourLatticeAveragingGroupFLAG:2021npn,
	archiveprefix = {arXiv},
	author = {Aoki, Y. and others},
	collaboration = {Flavour Lattice Averaging Group (FLAG)},
	doi = {10.1140/epjc/s10052-022-10536-1},
	eprint = {2111.09849},
	journal = {Eur. Phys. J. C},
	number = {10},
	pages = {869},
	primaryclass = {hep-lat},
	reportnumber = {CERN-TH-2021-191, JLAB-THY-21-3528, FERMILAB-PUB-21-620-SCD-T},
	title = {{FLAG Review 2021}},
	volume = {82},
	year = {2022}}

@article{MartinCamalich:2010fp,
	archiveprefix = {arXiv},
	author = {Martin Camalich, J. and Geng, L. S. and Vicente Vacas, M. J.},
	doi = {10.1103/PhysRevD.82.074504},
	eprint = {1003.1929},
	journal = {Phys. Rev. D},
	pages = {074504},
	primaryclass = {hep-lat},
	title = {{The lowest-lying baryon masses in covariant SU(3)-flavor chiral perturbation theory}},
	volume = {82},
	year = {2010}}

@article{Gegelia:1999gf,
	archiveprefix = {arXiv},
	author = {Gegelia, J. and Japaridze, G.},
	doi = {10.1103/PhysRevD.60.114038},
	eprint = {hep-ph/9908377},
	journal = {Phys. Rev. D},
	pages = {114038},
	title = {{Matching heavy particle approach to relativistic theory}},
	volume = {60},
	year = {1999}}

@article{Fuchs:2003qc,
	archiveprefix = {arXiv},
	author = {Fuchs, T. and Gegelia, J. and Japaridze, G. and Scherer, S.},
	doi = {10.1103/PhysRevD.68.056005},
	eprint = {hep-ph/0302117},
	journal = {Phys. Rev. D},
	pages = {056005},
	reportnumber = {MKPH-T-03-2},
	title = {{Renormalization of relativistic baryon chiral perturbation theory and power counting}},
	volume = {68},
	year = {2003}}

@article{Bar:2003mh,
	archiveprefix = {arXiv},
	author = {Bar, Oliver and Rupak, Gautam and Shoresh, Noam},
	doi = {10.1103/PhysRevD.70.034508},
	eprint = {hep-lat/0306021},
	journal = {Phys. Rev. D},
	pages = {034508},
	reportnumber = {UTHEP-469, LBNL-52989, BUHEP-03-13},
	title = {{Chiral perturbation theory at O(a**2) for lattice QCD}},
	volume = {70},
	year = {2004}}

@article{Agadjanov:2023efe,
	archiveprefix = {arXiv},
	author = {Agadjanov, Andria and Djukanovic, Dalibor and von Hippel, Georg and Meyer, Harvey B. and Ottnad, Konstantin and Wittig, Hartmut},
	doi = {10.1103/PhysRevLett.131.261902},
	eprint = {2303.08741},
	journal = {Phys. Rev. Lett.},
	number = {26},
	pages = {261902},
	primaryclass = {hep-lat},
	title = {{Nucleon Sigma Terms with Nf=2+1 Flavors of O(a)-Improved Wilson Fermions}},
	volume = {131},
	year = {2023}}

@article{Ren:2012aj,
	archiveprefix = {arXiv},
	author = {Ren, X.-L. and Geng, L.-S. and Martin Camalich, J. and Meng, J. and Toki, H.},
	doi = {10.1007/JHEP12(2012)073},
	eprint = {1209.3641},
	journal = {JHEP},
	pages = {073},
	primaryclass = {nucl-th},
	title = {{Octet baryon masses in next-to-next-to-next-to-leading order covariant baryon chiral perturbation theory}},
	volume = {12},
	year = {2012}}

@article{QCDSF-UKQCD:2003hmh,
	archiveprefix = {arXiv},
	author = {Ali Khan, A. and others},
	collaboration = {QCDSF-UKQCD},
	doi = {10.1016/j.nuclphysb.2004.04.018},
	eprint = {hep-lat/0312030},
	journal = {Nucl. Phys. B},
	pages = {175--194},
	reportnumber = {DESY-03-205, LTH-614, HU-EP-03-86, EDINBURGH-2003-24, LU-ITP-200-032},
	title = {{The Nucleon mass in N(f) = 2 lattice QCD: Finite size effects from chiral perturbation theory}},
	volume = {689},
	year = {2004}}

@article{Geng:2011wq,
	archiveprefix = {arXiv},
	author = {Geng, Li-Sheng and Ren, Xiu-Lei and Martin-Camalich, J. and Weise, W.},
	doi = {10.1103/PhysRevD.84.074024},
	eprint = {1108.2231},
	journal = {Phys. Rev. D},
	pages = {074024},
	primaryclass = {hep-ph},
	title = {{Finite-volume effects on octet-baryon masses in covariant baryon chiral perturbation theory}},
	volume = {84},
	year = {2011}}

@article{Symanzik:1983dc,
	author = {Symanzik, K.},
	doi = {10.1016/0550-3213(83)90468-6},
	journal = {Nucl. Phys. B},
	pages = {187--204},
	reportnumber = {DESY-83-016},
	title = {{Continuum Limit and Improved Action in Lattice Theories. 1. Principles and $\varphi^4$ Theory}},
	volume = {226},
	year = {1983}}

@article{Symanzik:1983gh,
	author = {Symanzik, K.},
	doi = {10.1016/0550-3213(83)90469-8},
	journal = {Nucl. Phys. B},
	pages = {205--227},
	reportnumber = {DESY-83-026},
	title = {{Continuum Limit and Improved Action in Lattice Theories. 2. O(N) Nonlinear Sigma Model in Perturbation Theory}},
	volume = {226},
	year = {1983}}

@article{Tiburzi:2005vy,
	archiveprefix = {arXiv},
	author = {Tiburzi, Brian C.},
	doi = {10.1016/j.nuclphysa.2005.07.011},
	eprint = {hep-lat/0501020},
	journal = {Nucl. Phys. A},
	pages = {232--258},
	title = {{Baryon masses at O(a**2) in chiral perturbation theory}},
	volume = {761},
	year = {2005}}

@article{Ren:2013wxa,
	archiveprefix = {arXiv},
	author = {Ren, Xiu-Lei and Geng, Li-Sheng and Meng, Jie},
	doi = {10.1140/epjc/s10052-014-2754-1},
	eprint = {1311.7234},
	journal = {Eur. Phys. J. C},
	number = {2},
	pages = {2754},
	primaryclass = {hep-ph},
	title = {{Baryon chiral perturbation theory with Wilson fermions up to $\mathcal{O}(a^2)$ and discretization effects of latest $n_f=2+1$ LQCD octet baryon masses}},
	volume = {74},
	year = {2014}}

\end{document}